\documentclass{article}

\usepackage{arxiv}

\usepackage[utf8]{inputenc} % allow utf-8 input
\usepackage[T1]{fontenc}    % use 8-bit T1 fonts
\usepackage{hyperref}       % hyperlinks
\usepackage{url}            % simple URL typesetting
\usepackage{booktabs}       % professional-quality tables
\usepackage{amsfonts}       % blackboard math symbols
\usepackage{nicefrac}       % compact symbols for 1/2, etc.
\usepackage{microtype}      % microtypography
\usepackage{graphicx}
\usepackage[numbers]{natbib}
\usepackage{doi}
\usepackage{xspace}

\newcommand*{\sqsn}{\ensuremath{\sqrt{s_{\rm NN}}}\xspace}

\newcommand*{\pT}{\ensuremath{p_\mathrm{T}}\xspace}
\newcommand*{\kT}{\ensuremath{k_\mathrm{T}}\xspace}

\newcommand*{\Raa}{\ensuremath{R_\mathrm{AA}}\xspace}
\newcommand*{\rg}{\ensuremath{r_\mathrm{g}}\xspace}
\newcommand*{\thg}{\ensuremath{\theta_\mathrm{g}}\xspace}

\title{Jet substructure measurements in heavy-ion collisions}

\date{May 26, 2024}	

\author{ \href{https://orcid.org/0000-0003-3706-5265}{\includegraphics[scale=0.06]{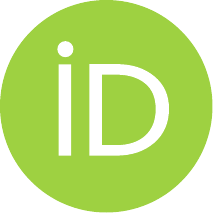}\hspace{1mm}Róbert Vértesi} \\ \\
	\textit{for the ATLAS, CMS, and ALICE Collaborations} \\ \\
	HUN-REN Wigner Research Centre for Physics,\\
	MTA Centre of Excellence, \\
	29-33 Konkoly-Thege Miklós út,\\
	1121 Budapest, Hungary\\
	\texttt{vertesi.robert@wigner.hu} \\
}

\renewcommand{\headeright}{}
\renewcommand{\undertitle}{}
\renewcommand{\shorttitle}{\textit{Jet substructure measurements in heavy-ion collisions}}

\hypersetup{
pdftitle={Jet substructure measurements in heavy-ion collisions},
pdfsubject={25.75.-q, 25.75.Nq, 25.75.Bh, 13.75.Cs},
pdfauthor={Robert Vertesi},
pdfkeywords={heavy-ion physics, high-energy physics, Large Hadron Collider, jets, jet substructure, fragmentation, quark--gluon plasma},
}

\begin{document}
\maketitle

\begin{abstract}
Jet substructure in heavy-ion collisions is a rapidly evolving area with lots of intriguing new measurements. This contribution presents a selection of recent jet-substructure measurements from experiments at the LHC, in particular, soft-drop groomed radii of jets and reclustered large-radius jets from ATLAS, jet axis difference and generalized jet angularities from ALICE, as well as dĳet shapes and b-jet shapes from the CMS experiment.
\end{abstract}

% keywords can be removed

\section{Introduction}

High-energy collisions of heavy ions produce hot and dense strongly coupled matter, the so-called quark--gluon plasma (QGP)~\cite{Shuryak:1980tp}. Jets, which are collimated particle sprays resulting from parton fragmentation, are modified by the QGP. The loss of jet momentum in the hot and dense medium, called jet quenching, was one of the first tell-tale signs of the presence of QGP~\cite{PHENIX:2001hpc}. 
Jet quenching alone, however, is not enough to fully understand the energy loss mechanisms of jets within the plasma. With systematic jet substructure measurements, however, not only can we investigate the energy loss of the color charge, but we can also learn about the length scales that are resolvable by the QGP, or determine whether the plasma has an emergent quasiparticle structure. 		
This paper summarizes some of the most interesting results of the ATLAS~\cite{ATLAS:2008xda}, CMS~\cite{CMS:2008xjf}, and ALICE~\cite{ALICE:2022wpn} experiments at the LHC, 
which contribute to answering these questions and extend our knowledge of the non-perturbative regime of quantum chromodynamics.

\section{Results}

\begin{figure}[h]
	\centering
	\begin{minipage}{.48\textwidth}
		\centering
		\includegraphics[width=\textwidth]{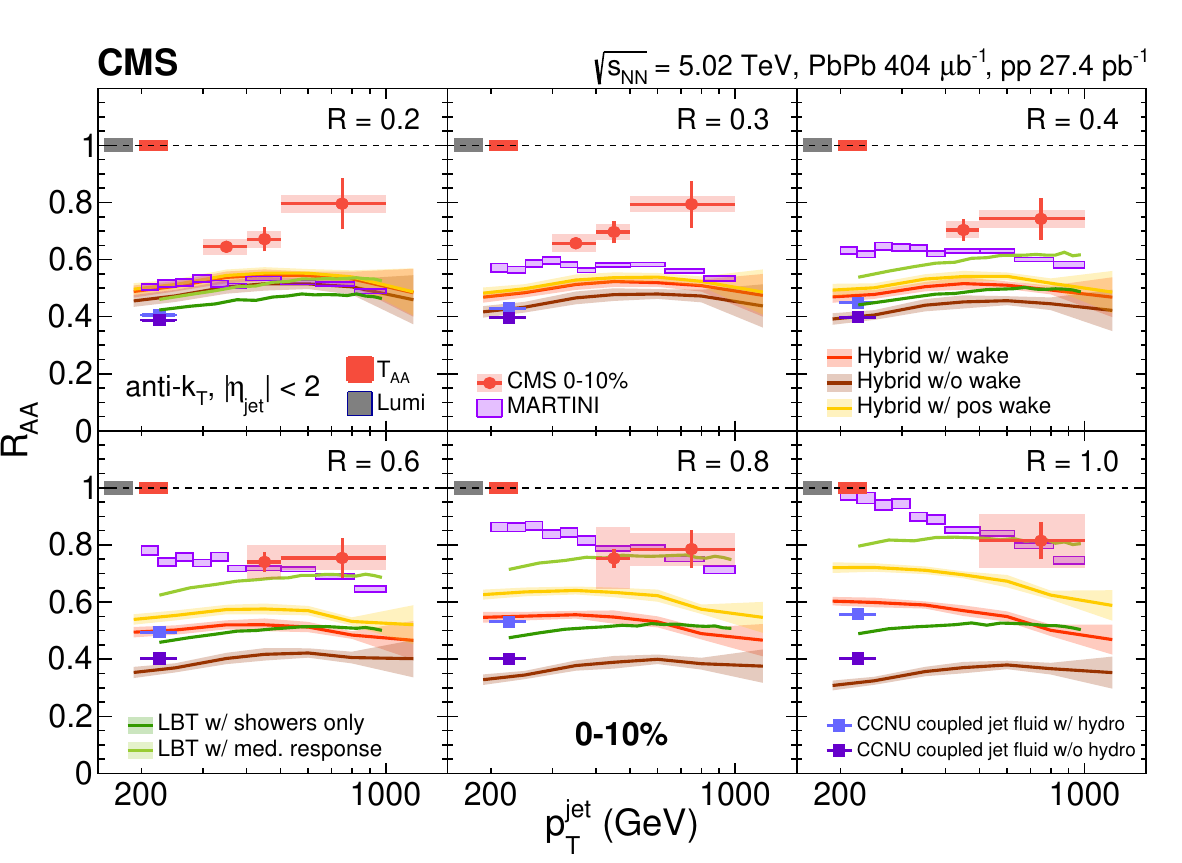}%
		\caption{
			Nuclear modification factor for jets reconstructed in the 0--10\% centrality class in Pb--Pb collisions at $\sqrt{s_{\rm NN}} = 5.02$ TeV.
			Measurements for different jet resolution parameters (R) are quoted in separate panels. Data are compared with predictions of several theoretical models~\cite{CMS:2021vui}.}
		\label{fig:CMS_Rdep}
	\end{minipage}%
	\hfill
	\begin{minipage}{.48\textwidth}
		\centering
		\includegraphics[width=\linewidth]{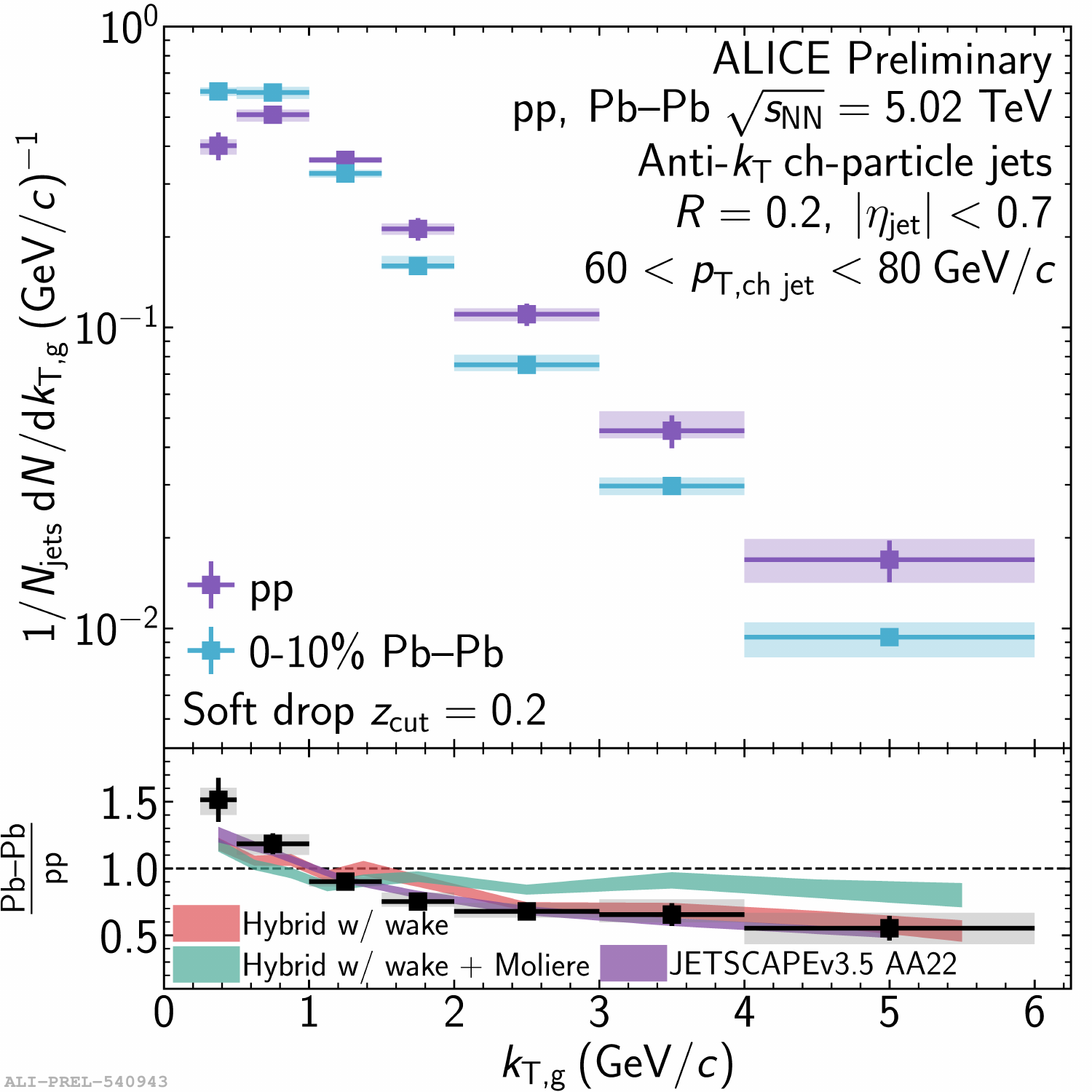}
		\caption{Comparison of the hardest $k_{\rm T,g}$ jet splitting found using SD $z_{\rm cut}=0.2$ for $R = 0.2$ jets in pp and 0--10\% PbPb collisions at $\sqsn = 5.02$ TeV.} 
		\label{fig:ALICE-kTg}
	\end{minipage}
\end{figure}

The dependence of the nuclear modification factor of jets on the jet resolution parameter exhibits a strong sensitivity to energy loss mechanisms as well as the medium response. Figure~\ref{fig:CMS_Rdep} presents the first measurement of large-radius jets in Pb--Pb collisions at $\sqrt{s_{\rm NN}} = 5.02$ TeV by the CMS experiment~\cite{CMS:2021vui}. A substantial suppression is visible at high momenta from small to large radii in central Pb--Pb collisions. The results also show a tension with models. Analysis of the jet substructure allows these mechanisms to be explored in detail.

In order to access the hard parton structure of the jet, so-called grooming algorithms are applied.
These are aimed at removing large-angle soft radiation and mitigate the effect of the underlying event and hadronization. Thus, groomed jets provide a direct interface to QCD calculations. 
In soft-drop (SD) grooming~\cite{Larkoski:2014wba}, jets that had previously been reconstructed with the anti-\kT algorithm~\cite{Cacciari:2008gp} are first declustered and then reclustered using the Cambridge--Aachen (C/A) algorithm~\cite{Dokshitzer:1997in} to form a clustering tree that follows angular ordering. Then the soft branches are iteratively removed if they do not fulfill the so-called soft-drop condition, 
\vspace{-2mm}
\begin{eqnarray}
	z > z_{\rm cut} \theta^\beta\,, & {\rm where} &  z = \frac{p_{{\rm T},2}}{p_{{\rm T},1}+p_{{\rm T},2}} \ {\rm and} \ \theta = \frac{\Delta R_{1,2} }{ R } 
\end{eqnarray}
are the momentum fraction taken by the subleading prong ($p_{{\rm T},1}$ and $p_{{\rm T},2}$ being the momenta of the two prongs), and the splitting radius (defined as the ratio of the $\Delta R_{1,2}$ splitting angle between the two prongs and the resolution parameter of the anti-\kT algorithm).
The parameter $z_{\rm cut}$ is the soft threshold. 
Rejection of soft wide-angle radiation can be also adjusted using the angular exponent $\beta$. The jet groomed momentum fraction $z_g$ and the groomed radius \thg, are defined as the values of $z$ and $\theta$ corresponding to the first hard splitting fulfilling the soft-drop condition. 

High relative-transverse-momentum (\kT) emissions can be a signature of point-like scattering. 
The ALICE experiment measured jets groomed with soft-drop ($z_{\rm cut} = 0.2$ and $\beta = 0$) as a function of the hardest-\kT splitting. The measured data are compared to model~\cite{Casalderrey-Solana:2014bpa,Putschke:2019yrg} in Fig.~\ref{fig:ALICE-kTg}. No clear enhancement can be seen at high \kT, and the model of Ref.~\cite{Casalderrey-Solana:2014bpa} describes the data better if there is no Molière scattering.

\begin{figure}[h]
	\centering
	\begin{minipage}{.48\textwidth}
		\centering
		\includegraphics[width=\linewidth]{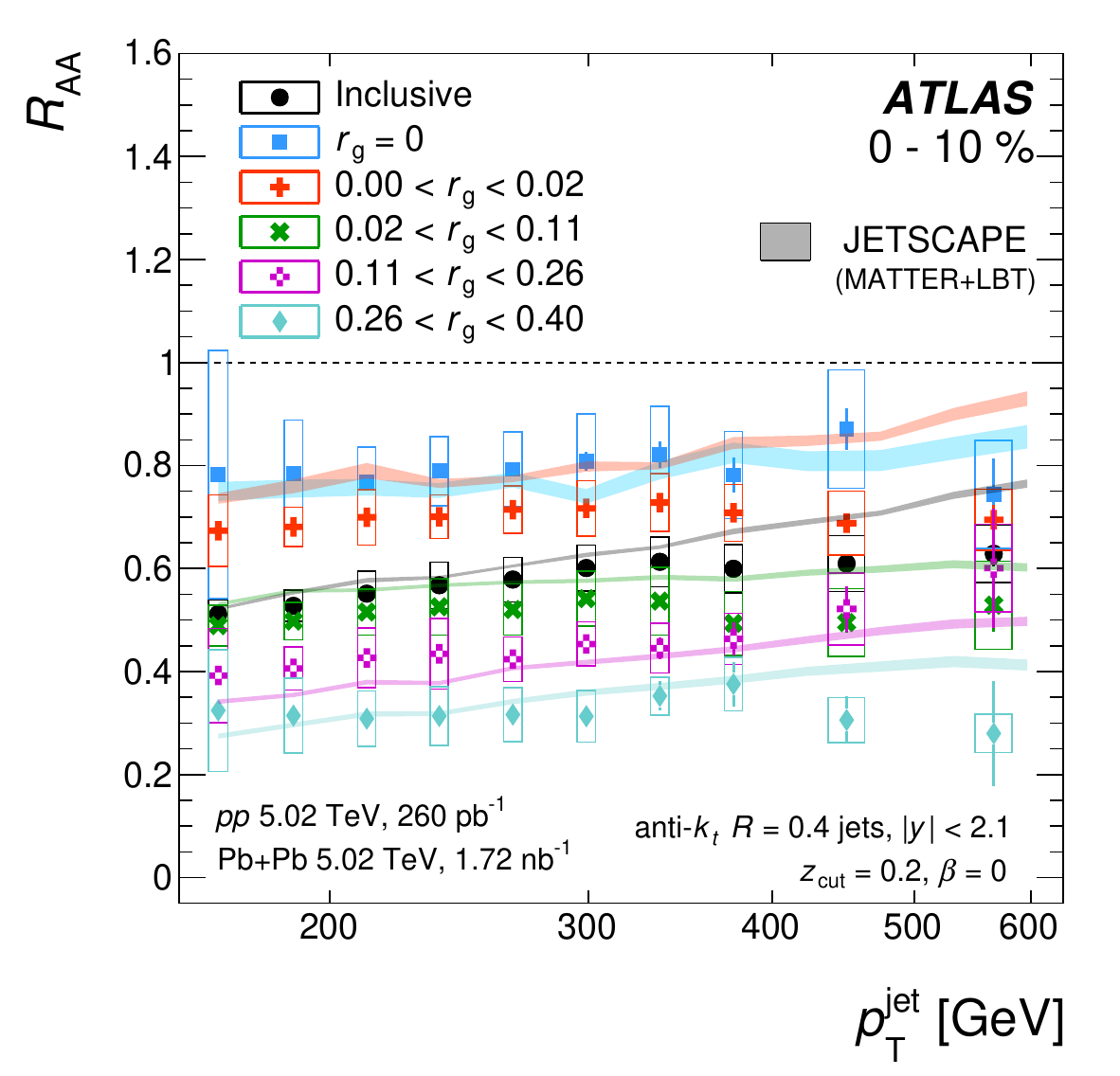}
		\caption{Comparison of \Raa as a function of jet \pT for inclusive jets and for four intervals of \rg~\cite{ATLAS:2022vii}.}
		\label{fig:ATLAS-Rg}
	\end{minipage}
	\hfill
	\begin{minipage}{.48\textwidth}
		\centering
		\includegraphics[width=\linewidth]{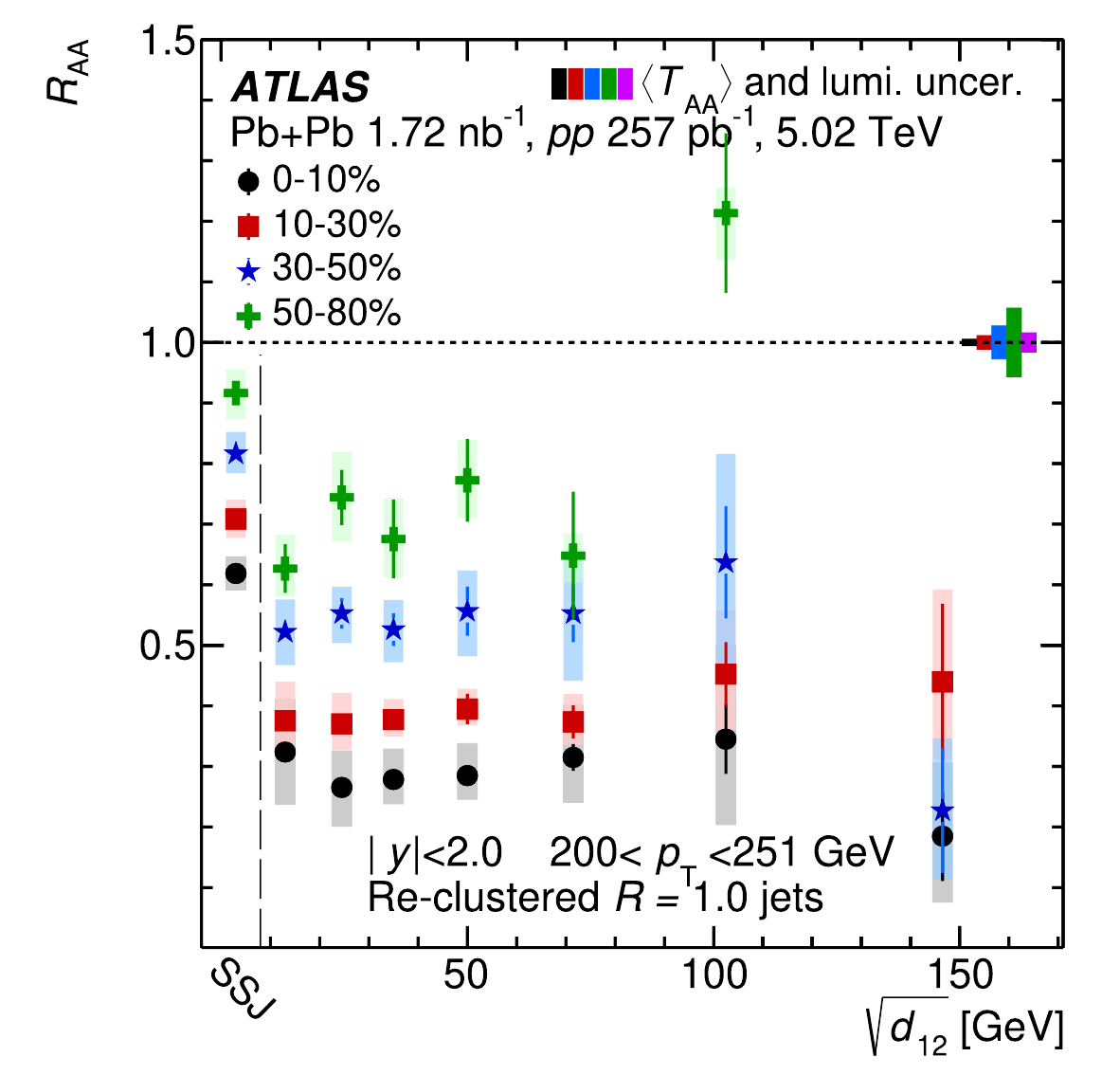}
		\caption{The values of \Raa for $R=1$ reclustered jets as a function of $\sqrt{d_{1,2}}$ in four centrality intervals~\cite{ATLAS:2023hso}.}
		\label{fig:ATLAS-subjet}
	\end{minipage}
\end{figure}

Jet energy loss is expected to depend on the jet substructure, since the extent to which the medium resolves the jet is connected to the decoherence in its radiation pattern. ATLAS measured the nuclear modification factor of jets with different $\rg = \thg R$ values, shown in Fig.~\ref{fig:ATLAS-Rg}~\cite{ATLAS:2022vii}.
Jets having a substructure with wider opening angle lose significantly more energy than jets with smaller \rg. The suppression does not depend strongly on \pT, regardless of \rg, which hints that the \pT-dependence of inclusive jet $R_{\rm AA}$ is driven by changes in \rg distribution. The results are qualitatively consistent with
a jet quenching model~\cite{Putschke:2019yrg}, which includes decoherence effects.

Jet substructures can be further explored using reclustered large-radius jets. In a new ATLAS measurement, small-radius ($R=0.2$) jets are reconstructed with the anti-\kT algorithm, and a $\pT>35$ GeV/$c$ threshold is applied. The remaining jets are then reconstructed into large-radius ($R=1.0$) jets using the \kT algorithm, and a splitting parameter is determined as
\begin{equation}
	\sqrt{d_{1,2}} = \min(p_{\rm T,1},p_{\rm T,1}) \Delta R_{1,2} \sim \kT \ ,
\end{equation} 
where $\Delta R_{1,2}$ is the angular separation and $p_{{\rm T},i}$ are the momenta of the two subjets. 
Figure~\ref{fig:ATLAS-subjet} shows the nuclear modification factor of these reclustered $R=1$ jets.
There is a significant difference in the quenching of large-radius jets having a single sub-jet and those with more complex substructure, where no pronunced dependence on $\sqrt{d_{1,2}}$ is seen. The observations support decoherence beyond a critical splitting angle~\cite{ATLAS:2023hso}. 

\begin{figure}[h]
	\centering
	\begin{minipage}{.48\textwidth}
		\centering
		\includegraphics[width=\linewidth]{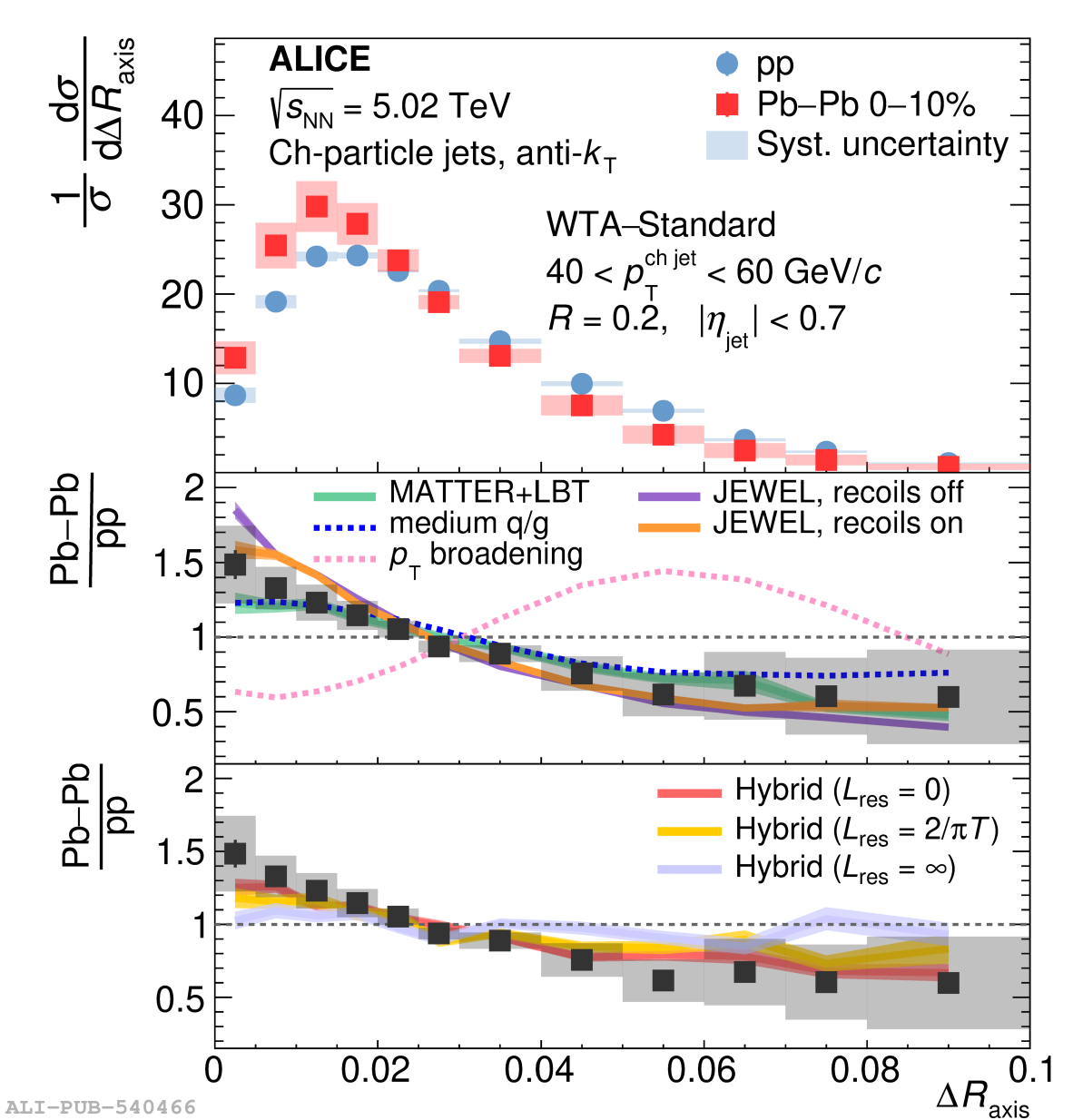}
		\caption{Central Pb--Pb and pp WTA-standard axis distributions (top), as well as the ratios compared to models (middle and bottom)~\cite{ALICE:2023dwg}.}
		\label{fig:ALICE-axis}
	\end{minipage}
	\hfill
	\begin{minipage}{.48\textwidth}
		\centering
		\includegraphics[width=\linewidth]{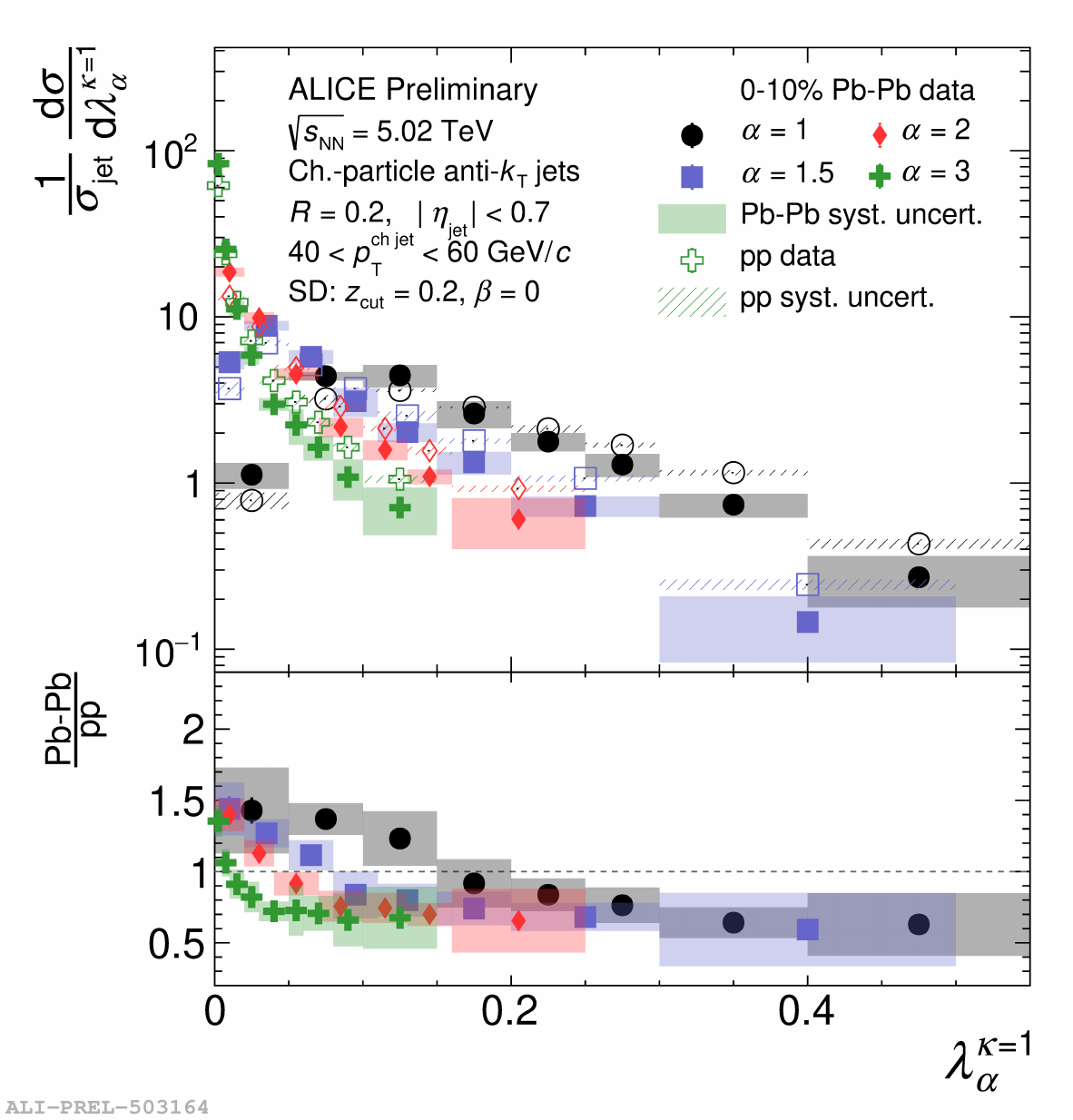}
		\caption{Groomed angularities of charged-particle $R=0.2$ jets with $40<\pT<60$ GeV/$c$ measured in Pb--Pb and pp~\cite{ALICE:2021njq} collisions at $\sqsn = 5.02$ TeV.}
		\label{fig:ALICE-ang}
	\end{minipage}
\end{figure}

ALICE measured the difference between the jet axes defined the following ways: {\it i)} standard axis, formed by the sum of pseudo-jet four-momenta in the clusterization with $E$-scheme; {\it ii)} soft-drop groomed jet axis, the sum of four-momenta of constituents accepted by the SD grooming; {\it iii)} winner-takes-all (WTA) axis, meaning that jets were reclustered with the C/A algorithm, where the prongs are always combined in the direction of the harder one, making the WTA axis insensitive to soft radiation~\cite{ALICE:2023dwg}. While standard--SD differences are relatively small, the WTA--standard axis results, shown in Fig.~\ref{fig:ALICE-axis}, are more substantial. The WTA--standard axis distribution exhibits a narrowing in heavy-ion collisions compared to pp collisions. The jet axis differences are also sensitive to the medium resolution length. From the comparison to the Hybrid model~\cite{Casalderrey-Solana:2014bpa}, the measurement favors incoherent energy loss. An intra-jet \pT broadening model~\cite{Ringer:2019rfk} does not describe the data trend. 

Generalized jet angularities are a class of observables that depend on both the longitudinal and angular properties of jet splittings,
\begin{eqnarray}
	\lambda^\kappa_\alpha = \sum_{i \in \rm{jet}} z^\kappa_i \theta^\alpha_i ,\ \ {\rm where}&
	z_i = \frac{p_{{\rm T},i}}{p_{\rm T,jet}} \ , & 
	\theta_i = \frac{\Delta R_{i,{\rm jet}}}{R} ,
\end{eqnarray} 
$p_{{\rm T},i}$ is the transverse momentum of the $i^{\rm th}$ component, and $\Delta R_{i,{\rm jet}}$ is its angle measured from the jet axis.
These are IRC-safe quantities for $\kappa = 1$ and $\alpha > 0$, therefore theoretically accessible in the vacuum case.
As they are a generalization of existing jet properties with continuously tunable parameters, $\lambda^1_1$ reduces to the jet girth and $\lambda^1_2$ to the jet thrust. 
Groomed and ungroomed generalized jet angularities reveal the effect of soft radiation.
In an ALICE measurement of groomed generalized jet angularities, shown in Fig.~\ref{fig:ALICE-ang}, a shift toward lower angularities can be observed, which suggests a narrowing of groomed jets in heavy-ion collisions. 	

\begin{figure}[h]
	\centering
	\begin{minipage}{.48\textwidth}
		\centering
		\includegraphics[width=\linewidth]{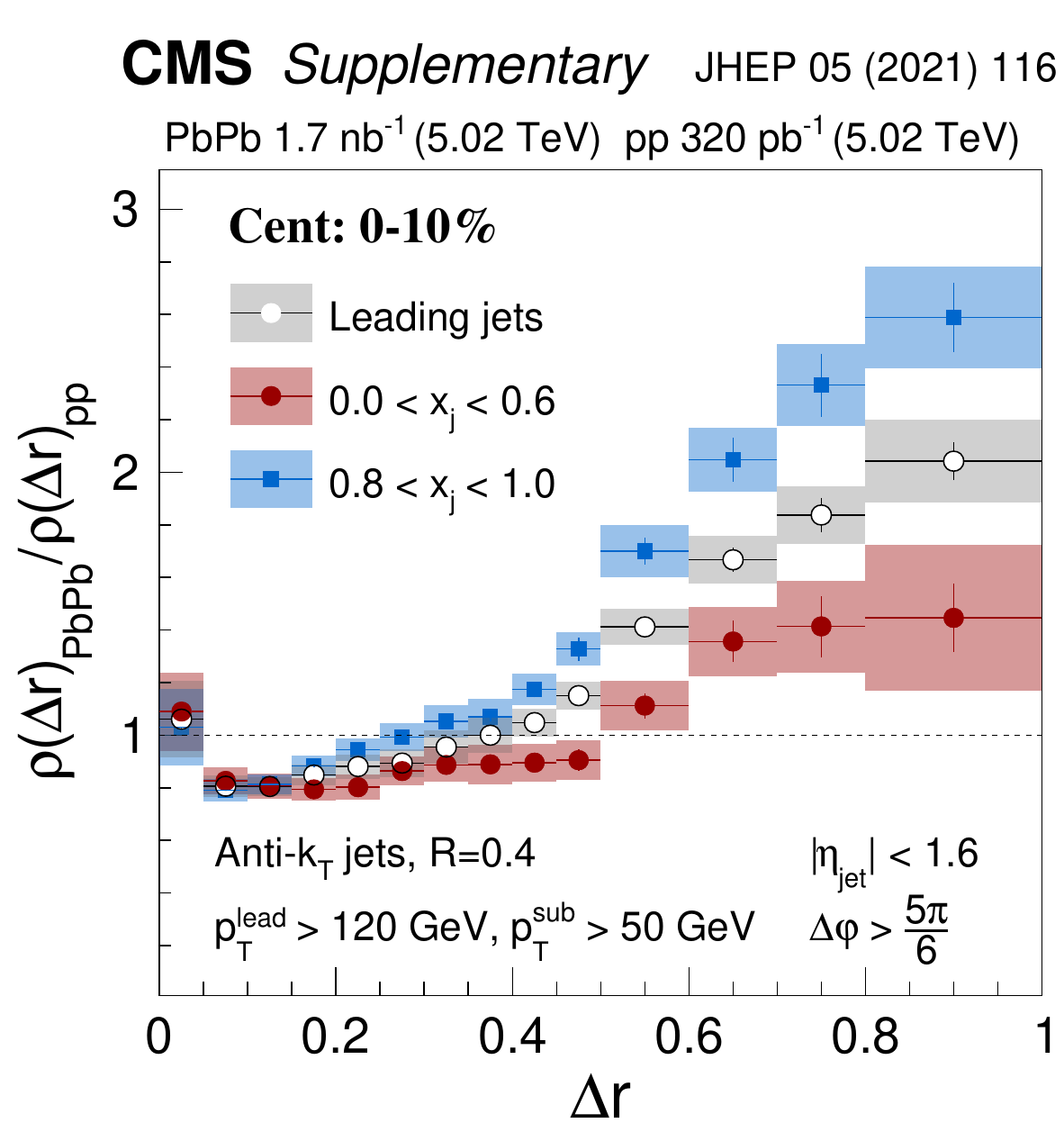}
		\caption{
			Pb--Pb to pp ratio for leading jet shapes for events in the 0--10\% centrality class, for different $x_j$ values~\cite{CMS:2021nhn}.}
		\label{fig:CMS-dijet}
	\end{minipage}
	\hfill
	\begin{minipage}{.48\textwidth}
		\centering
		\includegraphics[width=\linewidth]{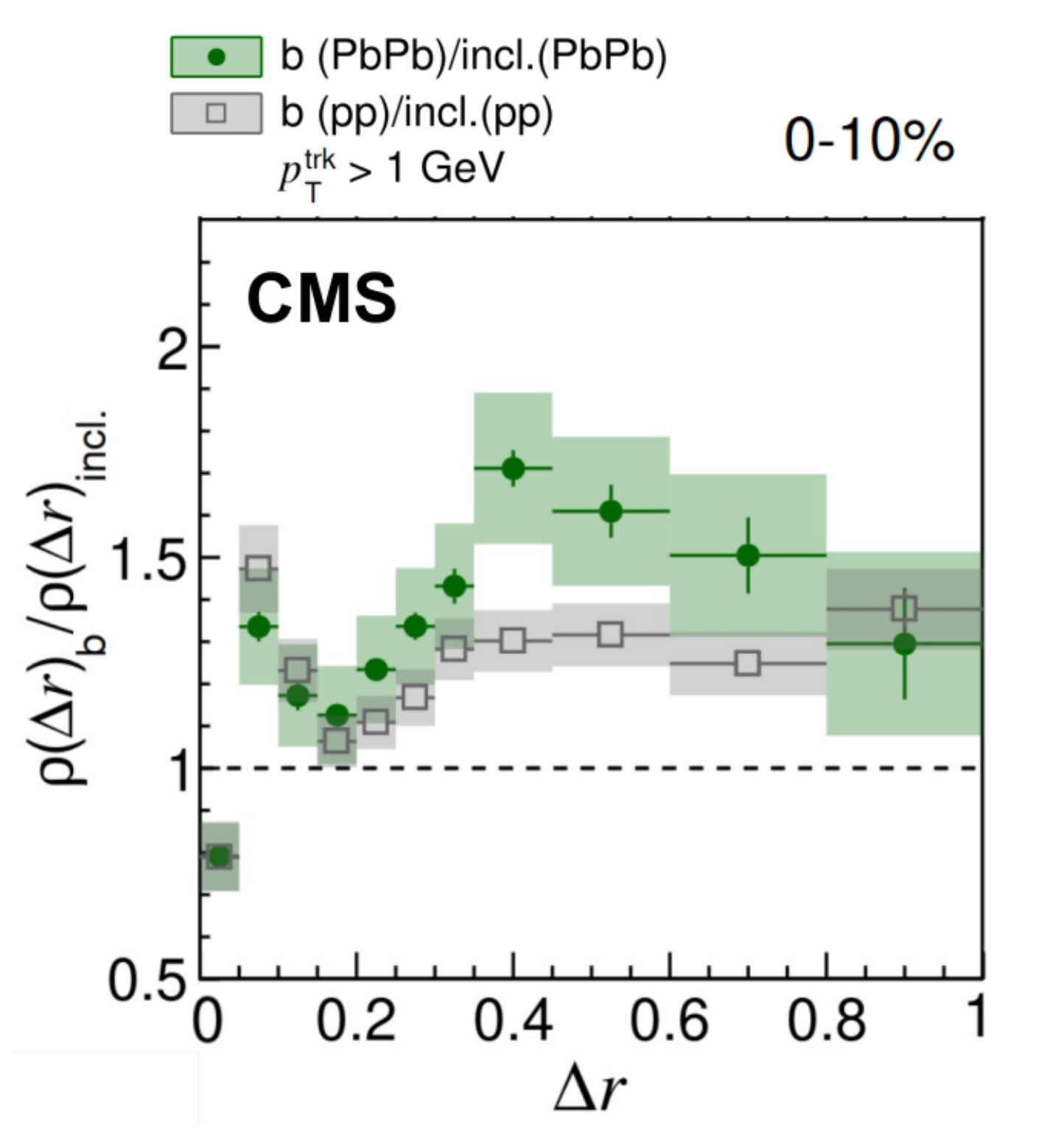}
		\caption{Ratio of b-jet to inclusive-jet shapes for 0--10\% centrality Pb--Pb collisions (green), as well as pp collisions (gray)~\cite{CMS:2022btc}.}
		\label{fig:CMS-bjet}
	\end{minipage}
\end{figure}

Jet shapes carry complementary information to groomed substructure measurements. However, they are sensitive to soft radiation, and background needs to be under control. 
In Ref.~\cite{CMS:2021nhn} CMS studied jet shape observable
\begin{equation}
	\rho(r) = \frac{1}{\delta r} \frac{1}{N_{\rm jets}}\sum_{\rm jets}\frac{\sum_{{\rm tracks} \in (r_a,r_b)}\pT^{\rm track}}{\pT^{\rm jet}} \ ,
\end{equation}
where $N_{\rm jets}$ is the number of jets, $r_a=r-\delta r/2$ and $r_b=r+\delta r/2$ are the radii corresponding to a ring of $\delta r$ width around the axis.
In this case, the jets were clustered with the anti-\kT algorithm using the WTA recombination scheme to suppress sensitivity of the jet shape to soft radiation.
Dijet shapes were further classified in terms of momentum imbalance between leading and subleading jets, $x_j = \pT^{\rm subleading} / \pT^{\rm leading}$. Figure~\ref{fig:CMS-dijet} shows the redistribution of energy from small angles with respect to the jet axis to larger angles, which is stronger for balanced jets. This hints on a path length dependence of the energy redistribution.

Figure~\ref{fig:CMS-bjet} shows the first measurement of b-jet shapes in high-energy collisions~\cite{CMS:2022btc}. 
In case of the Pb--Pb data, the ratio of jet shapes for  b-jets over inclusive jets exhibits an enhancement	with respect to the pp measurement. 
This indicates that medium response to the propagation of a heavier quark is stronger. 

\section{Summary}

We presented a selection of observables related to the jet substructure in high-energy heavy-ion collisions measured by the ATLAS, CMS, and ALICE collaborations. Only a tiny part of this rapidly  evolving area could be covered. While currently there is no clear evidence for point-like scattering centers, one can observe that the suppression of jets is strongly dependent on its substructure. 
Also, a general narrowing of the jet core can be observed. Pathlength-dependent modification patterns
are present, with an increased medium response to a heavier quark. 
With the advent of Run 3, increased sensitivity and new observables will be accessible such as 
energy--energy correlators, photon-tagged systems, and extended heavy-flavor measurements.

\section*{Acknowledgments}
The author is grateful for the discussions and review provided by colleagues from the ALICE, CMS and ATLAS collaborations. This work has been supported by the NKFIH OTKA FK131979 and 2021-4.1.2-NEMZ\_KI-2024-00034 projects. The author acknowledges the research infrastructure provided by the Hungarian Research Network (HUN-REN).

\bibliographystyle{unsrtnat}

\begin{thebibliography}{99}
	
	\bibitem{Shuryak:1980tp}
	E.~V.~Shuryak,
	``Quantum Chromodynamics and the Theory of Superdense Matter,''
	Phys. Rept. \textbf{61} (1980), 71
	%doi:10.1016/0370-1573(80)90105-2
	
	\bibitem{PHENIX:2001hpc}
	K.~Adcox \textit{et al.} [PHENIX],
	``Suppression of hadrons with large transverse momentum in central Au+Au collisions at $\sqrt{s_{NN}}$ = 130-GeV,''
	Phys. Rev. Lett. \textbf{88} (2002), 022301
	%doi:10.1103/PhysRevLett.88.022301
	[arXiv:nucl-ex/0109003 [nucl-ex]].
	
	\bibitem{ATLAS:2008xda}
	G.~Aad \textit{et al.} [ATLAS],
	``The ATLAS Experiment at the CERN Large Hadron Collider,''
	JINST \textbf{3} (2008), S08003.
	%doi:10.1088/1748-0221/3/08/S08003
	
	\bibitem{CMS:2008xjf}
	S.~Chatrchyan \textit{et al.} [CMS],
	``The CMS Experiment at the CERN LHC,''
	JINST \textbf{3} (2008), S08004.
	%doi:10.1088/1748-0221/3/08/S08004
	
	\bibitem{ALICE:2022wpn}
	ALICE Collaboration,
	``The ALICE experiment -- A journey through QCD,''
	arXiv:2211.04384 [nucl-ex].
	
	
	\bibitem{CMS:2021vui}
	A.~M.~Sirunyan \textit{et al.} [CMS],
	``First measurement of large area jet transverse momentum spectra in heavy-ion collisions,''
	JHEP \textbf{05} (2021), 284
	%doi:10.1007/JHEP05(2021)284
	[arXiv:2102.13080 [hep-ex]].
	
	\bibitem{Larkoski:2014wba}
	A.~J.~Larkoski, S.~Marzani, G.~Soyez and J.~Thaler,
	``Soft Drop,''
	JHEP \textbf{05} (2014), 146
	%doi:10.1007/JHEP05(2014)146
	[arXiv:1402.2657 [hep-ph]].
	
	%	\bibitem{Mehtar-Tani:2019rrk}
	%	Y.~Mehtar-Tani, A.~Soto-Ontoso and K.~Tywoniuk,
	%	%``Dynamical grooming of QCD jets,''
	%	Phys. Rev. D \textbf{101} (2020) no.3, 034004.
	%	%doi:10.1103/PhysRevD.101.034004
	%	%[arXiv:1911.00375 [hep-ph]].
	
	\bibitem{Cacciari:2008gp}
	M.~Cacciari, G.~P.~Salam and G.~Soyez,
	``The anti-$k_t$ jet clustering algorithm,''
	JHEP \textbf{04} (2008), 063
	%doi:10.1088/1126-6708/2008/04/063
	[arXiv:0802.1189 [hep-ph]].
	
	\bibitem{Dokshitzer:1997in}
	Y.~L.~Dokshitzer, G.~D.~Leder, S.~Moretti and B.~R.~Webber,
	``Better jet clustering algorithms,''
	JHEP \textbf{08} (1997), 001
	%doi:10.1088/1126-6708/1997/08/001
	[arXiv:hep-ph/9707323 [hep-ph]].
	
	\bibitem{Putschke:2019yrg}
	J.~H.~Putschke \textit{et al.}
	``The JETSCAPE framework,''
	[arXiv:1903.07706 [nucl-th]].
	
	\bibitem{Casalderrey-Solana:2014bpa}
	J.~Casalderrey-Solana, D.~C.~Gulhan, J.~G.~Milhano, D.~Pablos and K.~Rajagopal,
	``A Hybrid Strong/Weak Coupling Approach to Jet Quenching,''
	JHEP \textbf{10} (2014), 019; erratum: JHEP \textbf{09} (2015), 175
	%doi:10.1007/JHEP09(2015)175
	[arXiv:1405.3864 [hep-ph]].
	
	\bibitem{ATLAS:2022vii}
	G.~Aad \textit{et al.} [ATLAS],
	``Measurement of substructure-dependent jet suppr The International School for Advanced Studies (SISSA), find out more  The CMS experiment at the CERN LHCession in Pb+Pb collisions at 5.02 TeV with the ATLAS detector,''
	Phys. Rev. C \textbf{107} (2023) no.5, 054909.
	doi:10.1103/PhysRevC.107.054909
	[arXiv:2211.11470 [nucl-ex]].
	
	\bibitem{ATLAS:2023hso}
	G.~Aad \textit{et al.} [ATLAS],
	``Measurement of Suppression of Large-Radius Jets and Its Dependence on Substructure in Pb+Pb Collisions at sNN=5.02\,\,TeV with the ATLAS Detector,''
	Phys. Rev. Lett. \textbf{131} (2023) no.17, 172301
	%doi:10.1103/PhysRevLett.131.172301
	[arXiv:2301.05606 [nucl-ex]].
	
	\bibitem{ALICE:2023dwg}
	S.~Acharya \textit{et al.} [ALICE],
	``Measurement of the angle between jet axes in Pb$-$Pb collisions at $\sqrt{s_{\rm NN}} = 5.02$ TeV,''
	arXiv:2303.13347 [nucl-ex].
	
	\bibitem{ALICE:2021njq}
	S.~Acharya \textit{et al.} [ALICE],
	``Measurements of the groomed and ungroomed jet angularities in pp collisions at $ \sqrt{s} $ = 5.02 TeV,''
	JHEP \textbf{05} (2022), 061
	%doi:10.1007/JHEP05(2022)061
	[arXiv:2107.11303 [nucl-ex]].
	
	\bibitem{Ringer:2019rfk}
	F.~Ringer, B.~W.~Xiao and F.~Yuan,
	``Can we observe jet $P_T$-broadening in heavy-ion collisions at the LHC?,''
	Phys. Lett. B \textbf{808} (2020), 135634
	%doi:10.1016/j.physletb.2020.135634
	[arXiv:1907.12541 [hep-ph]].
	
	%	\bibitem{Kang:2018qra}
	%	Z.~B.~Kang, K.~Lee and F.~Ringer,
	%	%``Jet angularity measurements for single inclusive jet production,''
	%	JHEP \textbf{04} (2018), 110.
	%	%doi:10.1007/JHEP04(2018)110
	%	%[arXiv:1801.00790 [hep-ph]].
	
	\bibitem{CMS:2021nhn}
	A.~M.~Sirunyan \textit{et al.} [CMS],
	``In-medium modification of dijets in PbPb collisions at $ \sqrt{s_{\mathrm{NN}}} $ = 5.02 TeV,''
	JHEP \textbf{05} (2021), 116
	%doi:10.1007/JHEP05(2021)116 The International School for Advanced Studies (SISSA), find out more  The CMS experiment at the CERN LHC
	[arXiv:2101.04720 [hep-ex]].
	
	\bibitem{CMS:2022btc}
	A.~Tumasyan \textit{et al.} [CMS],
	``Search for medium effects using jets from bottom quarks in PbPb collisions at $\sqrt{s_\mathrm{NN}}$ = 5.02 TeV,''
	Phys. Lett. B \textbf{844} (2023), 137849
	%doi:10.1016/j.physletb.2023.137849
	[arXiv:2210.08547 [hep-ex]].
	
\end{thebibliography}

\end{document}